\documentclass[a4paper,11pt]{article}
\pdfoutput=1 

\usepackage{jinstpub}
\usepackage[squaren,Gray,cdot]{SIunits}
\pdfmapfile{newtx.map}

\title{\boldmath R\&D towards the CMS RPC Phase-2 upgrade}

\author[a,1]{A. Fagot\note{Corresponding author.}}
\author[a]{, A. Cimmino, S. Crucy, M. Gul, A.A.O. Rios, M. Tytgat, N. Zaganidis}
\author[b]{, S. Aly, Y. Assran, A. Radi, A. Sayed}
\author[c]{, G. Singh}
\author[d]{, M. Abbrescia, G. Iaselli, M. Maggi, G. Pugliese, P. Verwilligen}
\author[e]{, W. Van Doninck}
\author[f]{, S. Colafranceschi, A. Sharma}
\author[g]{, L. Benussi, S. Bianco, D.Piccolo, F. Primavera}
\author[h]{, V. Bhatnagar, R. Kumari, A. Mehta, J. Singh}
\author[i]{, A. Ahmad, W. Ahmed, M. I. Asghar, I. M. Awan, H. R. Hoorani, S. Muhammad, M.A. Shah, H. Shahzad}
\author[j]{, S.W. Cho, S.Y. Choi, B. Hong, M.H. Kang, K.S. Lee, J.H. Lim, S.K. Park}
\author[k]{, M.S. Kim}
\author[l]{, M. Gouzevitch, G. Grenier, F. Lagarde, I.B. Laktineh}
\author[m]{, S. Carpinteyro Bernardino, C. Uribe Estrada, I. Pedraza}
\author[n]{, S. Carrillo Moreno, F. Vazquez Valencia}
\author[o]{, L.M. Pant}
\author[p]{, S. Buontempo, N. Cavallo, M. Esposito, F. Fabozzi, G. Lanza, I. Orso, L. Lista, S. Meola, M. Merola, P. Paolucci, F. Thyssen}
\author[q]{, A. Braghieri, A. Magnani, P. Montagna, C. Riccardi, P. Salvini, I. Vai, P. Vitulo}
\author[r]{, Y. Ban, S.J. Qian}
\author[s]{, M. Choi}
\author[t]{, Y. Choi, J. Goh, D. Kim}
\author[u]{, A. Aleksandrov, R. Hadjiiska, P. Iaydjiev, M. Rodozov, S. Stoykova, G. Sultanov, M. Vutova}
\author[v]{, A. Dimitrov, L. Litov, B. Pavlov, P. Petkov}
\author[w]{, I. Bagaturia, D. Lomidze}
\author[x]{, C. Avila, A. Cabrera, J.C. Sanabria}
\author[y]{, I. Crotty}
\author[z]{, J. Vaitkus}

\affiliation[a]{Ghent university, Dept. of Physics and Astronomy, Proeftuinstraat 86, B-9000 Ghent, Belgium}
\affiliation[b]{Egyptian Network for High Energy Physics, Academy of Scientific Research and Technology, 101 Kasr El-Einy St. Cairo Egypt.}
\affiliation[c]{Chulalongkorn University, Department of Physics, Faculty of Science, Payathai Road, Phatumwan, Bangkok, THAILAND - 10330.}
\affiliation[d]{INFN, Sezione di Bari, Via Orabona 4, IT-70126 Bari, Italy.}
\affiliation[e]{Vrije Universiteit Brussel, Boulevard de la Plaine 2, 1050 Ixelles, Belgium.}
\affiliation[f]{Physics Department CERN, CH-1211 Geneva 23, Switzerland.}
\affiliation[g]{INFN, Laboratori Nazionali di Frascati (LNF), Via Enrico Fermi 40, IT-00044 Frascati, Italy.}
\affiliation[h]{Department of Physics, Panjab University, Chandigarh Mandir 160 014, India.}
\affiliation[i]{National Centre for Physics, Quaid-i-Azam University, Islamabad, Pakistan.}
\affiliation[j]{Korea University, Department of Physics, 145 Anam-ro, Seongbuk-gu, Seoul 02841, Republic of Korea.}
\affiliation[k]{Kyungpook National University, 80 Daehak-ro, Buk-gu, Daegu 41566, Republic of Korea.}
\affiliation[l]{Universite de Lyon, Universite Claude Bernard Lyon 1, CNRS-IN2P3, Institut de Physique Nucleaire de Lyon, Villeurbanne, France.}
\affiliation[m]{Benemerita Universidad Autonoma de Puebla, Puebla, Mexico.}
\affiliation[n]{Universidad Iberoamericana, Mexico City, Mexico.}
\affiliation[o]{Nuclear Physics Division Bhabha Atomic Research Centre Mumbai 400 085, INDIA.}
\affiliation[p]{INFN, Sezione di Napoli, Complesso Univ. Monte S. Angelo, Via Cintia, IT-80126 Napoli, Italy.}
\affiliation[q]{INFN, Sezione di Pavia, Via Bassi 6, IT-Pavia, Italy.}
\affiliation[r]{School of Physics, Peking University, Beijing 100871, China.}
\affiliation[s]{University of Seoul, 163 Seoulsiripdae-ro, Dongdaemun-gu, Seoul, Republic of Korea.}
\affiliation[t]{Sungkyunkwan University, 2066 Seobu-ro, Jangan-gu, Suwon-si, Gyeonggi-do, Republic of Korea.}
\affiliation[u]{Bulgarian Academy of Sciences, Inst. for Nucl. Res. and Nucl. Energy, Tzarigradsko shaussee Boulevard 72, BG-1784 Sofia, Bulgaria.}
\affiliation[v]{Faculty of Physics, University of Sofia,5, James Bourchier Boulevard, BG-1164 Sofia, Bulgaria.}
\affiliation[w]{Tbilisi University, 1 Ilia Chavchavadze Ave, Tbilisi 0179, Georgia.}
\affiliation[x]{Universidad de Los Andes, Apartado Aereo 4976, Carrera 1E, no. 18A 10, CO-Bogota, Colombia.}
\affiliation[y]{Dept. of Physics, Wisconsin University, Madison, WI 53706, United States.}
\affiliation[z]{Vilnius University, Vilnius, Lithuania.}

\emailAdd{alexis.fagot@ugent.be}

\abstract{The high pseudo-rapidity region of the CMS muon system is covered by Cathode Strip Chambers (CSC) only and lacks redundant coverage despite the fact that it is a challenging region for muons in terms of backgrounds and momentum resolution. In order to maintain good efficiency for the muon trigger in this region additional RPCs are planned to be installed in the two outermost stations at low angle named RE3/1 and RE4/1. These stations will use RPCs with finer granularity and good timing resolution to mitigate background effects and to increase the redundancy of the system.}


\arxivnumber{}

\proceeding{13$^{\text{th}}$ Workshop on Resistive Plate Chambers and related detectors\\
  22-26 February 2016\\
  Ghent University, Belgium}

\begin{document}
\maketitle
\flushbottom

\section{Introduction}
\label{sec:intro}

After the more than two years lasting first Long Shutdown (LS1), the Large Hadron Collider (LHC) delivered its very first Run-II proton-proton collisions early 2015. LS1 gave the opportunity to the LHC and to the its experiments to undergo upgrades. The accelerator is now providing collisions at center-of-mass energy of \unit{13}{TeV} and bunch crossing rate of \unit{40}{MHz}, with a peak luminosity exceeding its design value. During the first and upcoming second LHC Long Shutdown, the Compact Muon Solenoid (CMS) detector is also undergoing a number of upgrades to maintain a high system performance~\cite{MUONTDR}.

\begin{figure}[ht!]
	\centering
	\includegraphics[width=0.7\textwidth]{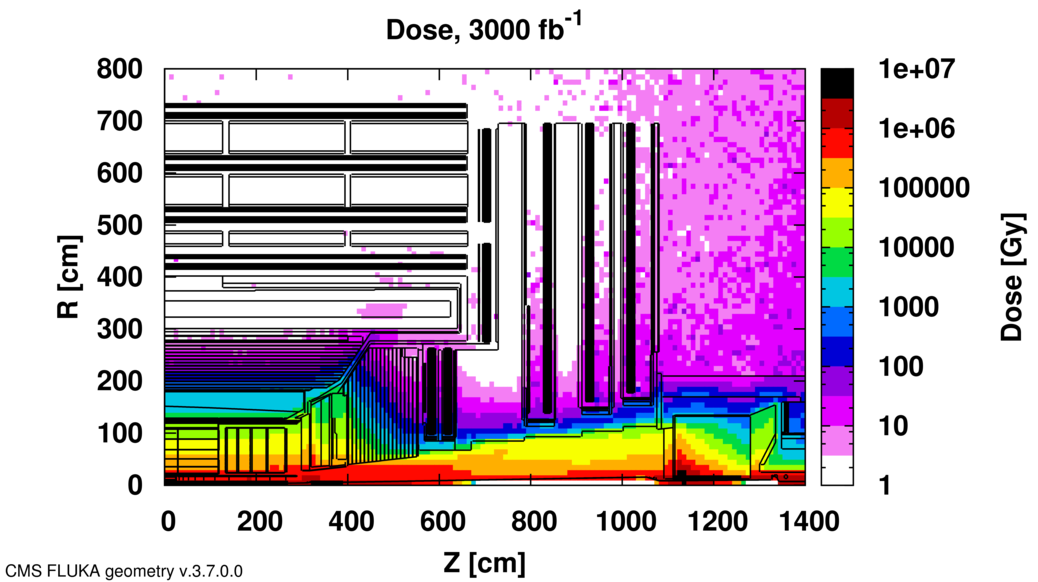}
	\caption{\label{fig:Dose} Absorbed dose in the CMS cavern after an integrated luminosity of \unit{3000}{\femto\reciprocal\barn}. R is the transverse distance from the beamline and Z is the distance along the beamline from the Interaction Point at Z=0.}
\end{figure}

From the LHC Phase-2 or High-Luminosity (HL-LHC) period onwards, i.e. past the third LHC Long Shutdown (LS3), the performance degradation due to integrated radiation as well as the average number of inelastic collisions per bunch crossing, or pileup, will rise substantially and become a major challenge for the LHC experiments, like CMS that are forced to address an upgrade program for Phase-II~\cite{PHASEIITP}. Simulations of the expected distribution of absorbed dose in the CMS detector under HL-LHC conditions, show in figure~\ref{fig:Dose} that detectors placed close to the beamline will have to withstand high irradiation, the radiation dose being of the order of a few tens of\unit{}{Gy}.

\begin{figure}[ht!]
	\centering
	\includegraphics[width=0.7\textwidth]{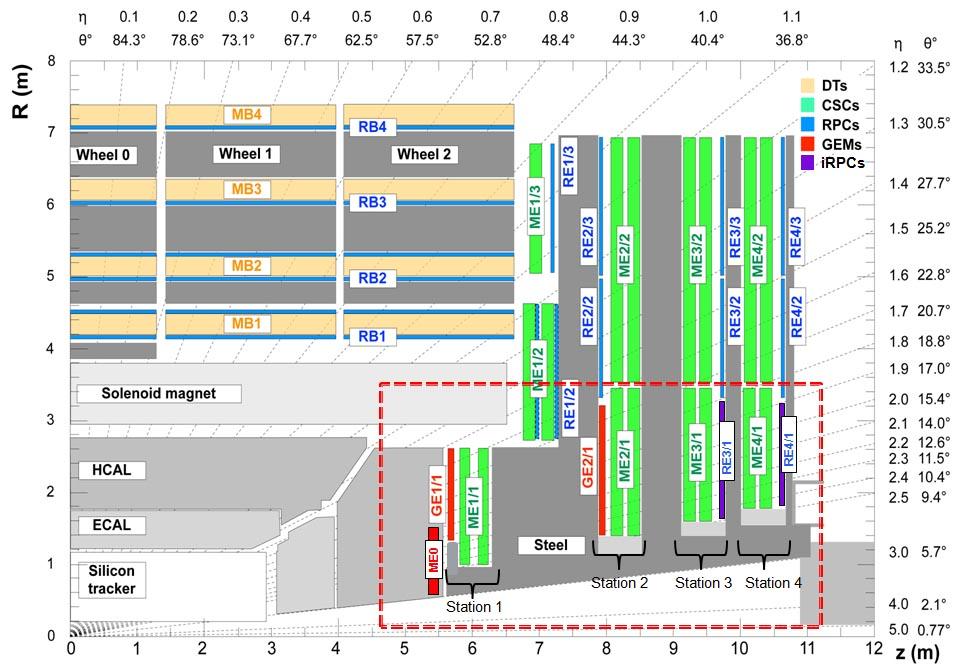}
	\caption{\label{fig:Quadrant} A quadrant of the muon system, showing DT chambers (yellow), RPC (light blue), and CSC (green). The locations of new forward muon detectors for Phase-II are contained within the dashed box and indicated in red for GEM stations (ME0, GE1/1, and GE2/1) and dark blue for improved RPC stations (RE3/1 and RE4/1).}
\end{figure}

The measurement of small production cross-section and/or decay branching ratio processes, such as the Higgs boson coupling to charge leptons or the $B_s \longrightarrow \mu^+\mu^-$ decay, is of major interest and specific upgrades in the forward regions of the detector will be required to maximize the physics acceptance on the largest possible solid angle. To ensure proper trigger performance within the present coverage, the muon system will be completed with new chambers. In figure~\ref{fig:Quadrant} one can see that the existing Cathode Strip Chamber (CSC) modules will be completed by Gas Electron Multipliers (GEM) and Resistive Plate Chambers (RPC) in the pseudo-rapidity region $1.6<\vert\eta\vert<2.4$ to complete its redundancy as originally scheduled in the CMS Technical Proposal~\cite{CMSTP}.

\begin{figure}[ht!]
	\centering
	\includegraphics[width=0.6\textwidth]{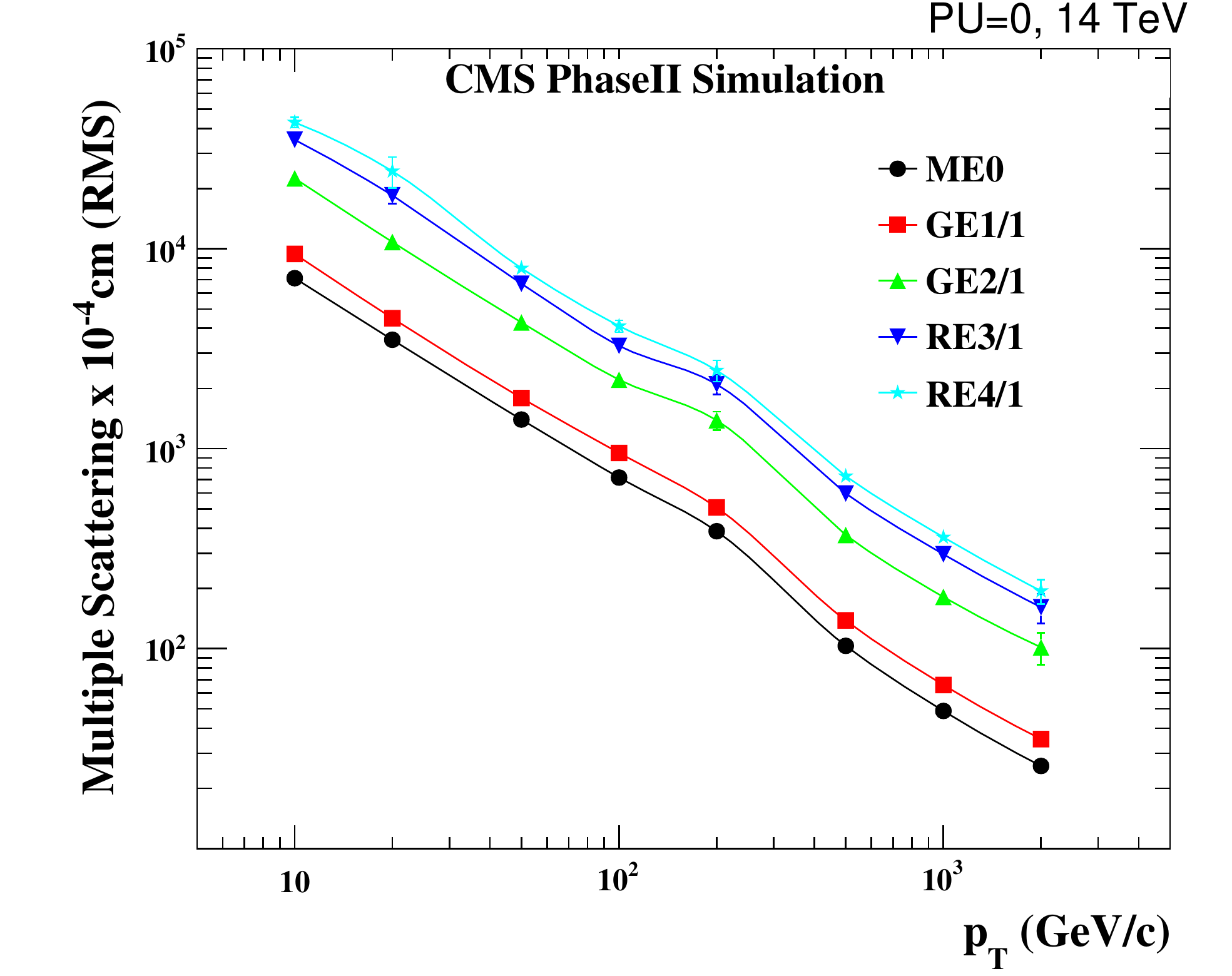}
	\caption{\label{fig:MultiScat}  RMS of the multiple scattering displacement as a function of muon $p_T$ for the  proposed forward muon stations. All of the electromagnetic processes such as bremsstrahlung and magnetic field effect are included in the simulation.}
\end{figure}

RPCs are used by the CMS first level trigger for their good timing performances. Indeed, a very good bunch crossing identification can be obtained with the present CMS RPC system, given their fast response of the order of \unit{1}{ns}. In order to contribute to the precision of muon momentum measurements, muon chambers should have a spatial resolution less or comparable to the contribution of multiple scattering~\cite{MUONTDR}. Most of the plausible physics is covered only considering muons with $p_T<$\unit{100}{GeV} thus, in order to match CMS requirements, a spatial resolution of $\mathcal{O}$(few $\mathrm{mm}$) the proposed new RPC stations, as shown by the simulation in figure~\ref{fig:MultiScat}. According to preliminary designs, RE3/1 and RE4/1 readout pitch will be comprised between 3 and \unit{6}{mm} and 5 $\eta$-partitions could be considered.

\section{Irradiation tests at the CERN GIF++}
\label{sec:GIF++}

\begin{figure}[ht!]
	\centering
	\includegraphics[width=0.7\textwidth]{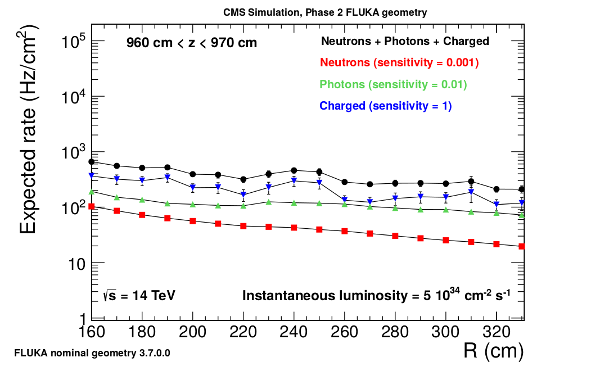}
	\caption{\label{fig:Rate} Simulated particle hit rate as a function of radius at the RE3/1 station, assuming a 0.001 sensitivity to neutrons, 0.01 to photons and 1 to charged, ionizing particles.}
\end{figure}

In the very forward region of CMS, i.e. close to the LHC beam line, the third endcap disk will be subjected to a total expected background rate of the order of \unit{0.6}{kHz/cm\squared} according to figure~\ref{fig:Rate} requiring solutions for the new RPC stations with stable, high performance in time in a high radiation environment for the new RE3/1 and RE4/1 stations~\cite{PHASEIITP}. Such rate is equivalent to an integrated charge of approximately \unit{1}{C/cm\squared} during the lifetime of the detectors assuming a mean charge deposition $\langle q \rangle$ per avalanche of \unit{20}{pC} and a safety factor of 2.

To ensure their robustness, CMS present muon system detectors and new R\&D efforts will be irradiated and monitored. The GIF++ at CERN is a new gamma irradiation facility to test detectors for the HL-LHC program. High energy $\mu$ beam with momentum up to \unit{100}{GeV/c} is provided and combined with a \unit{14}{TBq} $^{137}$Cs source. Performance of RPCs will be thoroughly tested with high radiation background. At HL-LHC, to integrate \unit{3000}{\femto\reciprocal\barn} at a luminosity $L = 5\times 10^{34}$\unit{}{\centi\rpsquare\meter\cdot\reciprocal\second}, an effective time $T_{eff} = 6\times 10^7$\unit{}{s} is needed.

\begin{eqnarray}
	Q^{HL-LHC}_{int} & = & \langle q \rangle \cdot T_{eff} \cdot \Phi_{eff} \label{eq1}\\
	Q^{GIF}_{int} & = & \langle q \rangle \cdot T_{irr} \cdot \Phi_{eff} \cdot AF \label{eq2}
\end{eqnarray}

Using equations~\ref{eq1} and~\ref{eq2}, and an acceleration factor $AF = 2$, an irradiation time $T_{irr}$ of approximately 17 months will be necessary to reproduce a charge deposition similar to HL-LHC. RPC life time is dependant on the total integrated charge over time that can degrade the material and components inside the detector, and contribute to produce HF~\cite{HF}. Test in GIF++ will help us understand the HF production through time in RPCs using fluoride-rich gas mixture as well as their overall ageing.

\section{Investigated forward RPC technologies}
\label{sec:technologies}

As RPCs are resistive detectors, increasing the rate capability of the new generation via a reduction of their electrode's resistivity is a first possibility to be investigated in order to accelerate the charge recombination at the level of the electrode's surface.

A reduction of resistivity should be coupled with a reduction of mean charge deposition per avalanche. Equipping RPCs with more sensitive front-end electronics and moving a part of the amplification to the electronics would allow to work at lower voltages. Another solution would be to improve the current CMS RPC design. For example, changing the number of electrodes or their thickness could yield a better ratio of induced signal to moving charge.

By combining these different techniques, a reduction of the voltage drop over the electrodes is expected with improvements on both rate capability and ageing of the detector.

\subsection{Electrode materials}
\label{subsec:electrodes}

Materials that are being used in RPCs are listed into table~\ref{tab:electrodes}. The materials presented here are the Bakelite and Low Resistive Silicate (LRS) glass considered by CMS together with other doped glass and ceramics. HPL being industrially produced offers lower costs and bigger material surfaces allowing to build wide area detectors, whereas glass and ceramics are still prototype materials produced locally at higher costs and small surfaces. Glass and ceramics have smooth surfaces, naturally offering uniform electric fields, and offer a tunable range of resistivity values, while Bakelite has only limited resistivity range and with its porous surface usually requires a linseed oil treatment.

\begin{table}
	\centering
	\begin{tabular}{|c|c|c|}
		\hline
		Electrode material & $\rho$(\unit{}{\ohm\centi\meter}) & Institutes \\
		\hline
		\hline
		HPL & $0.5-1\times 10^{10}$ & INFN\cite{KODEL} \\
		\hline
		LRS glass & $10^{10}$ & IPNL-LLR-Tsinghua\cite{YACINE,TSINGHUA} \\
		\hline
		Vanadate glass & $10^{4}$ to $10^{16}$ & Coe College-ANL-University of Iowa\cite{JOSE} \\
		\hline
		SiC based ceramics & $10^{7}$ to $10^{12}$ & HZDR\cite{CERAMICS} \\
		\hline
		Ferrite ceramics &  $10^{6}$ to $10^{13}$ & CSIC-USC\cite{MIGUEL} \\
		\hline
	\end{tabular}
	\caption{\label{tab:electrodes} List of materials used for improved RPC prototypes. A resistivity range is given for Vanadate glass, SiC based ceramics and Ferrite ceramics for which the production process allows a fine tuning of the resistivity.}
\end{table}

\subsection{Front-end electronics}
\label{subsec:electronics}

Only changing the electrode resistivity to make RPCs faster won't help in reducing the ageing effects to ensure smooth performance over a long time period. A main contribution to the ageing is due to the amount of deposited charge per avalanche created in the gas gap. Being able to reduce this mean charge deposition will lead to a reduction the total integrated charge over long periods. This can be achieved via more sensitive low-noise front-end electronics.

\begin{figure}[ht!]
	\centering
	\includegraphics[width=0.7\textwidth]{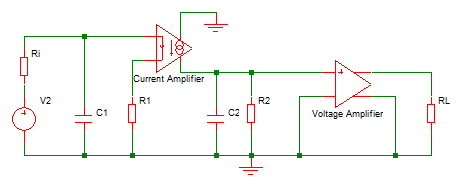}
	\qquad
	\includegraphics[width=0.7\textwidth]{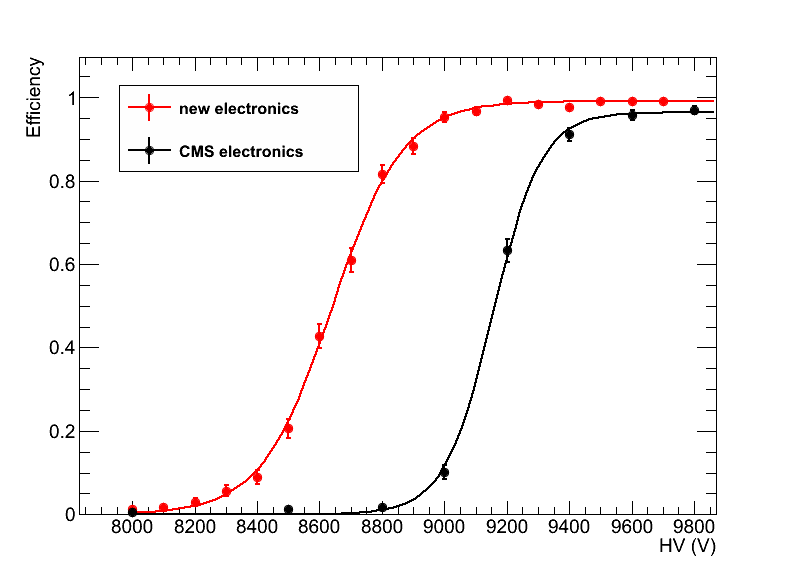}
	\caption{\label{fig:ATLAS} Left: block diagram of INFN amplifier where R1 is the input impedance resistance. C1$\ll$C2 and R1$\ll$R2 are the conditions for the circuit to work properly. Right: efficiency plateaus for the cases of the standard CMS RPC front-end electronics and new INFN prototype electronics. A lower charge crossing the RPC gap is indicated by the shift of the plateau toward lower voltages leading to an improved rate capability of the RPC.}
\end{figure}

\begin{figure}[ht!]
	\centering
	\includegraphics[width=0.7\textwidth]{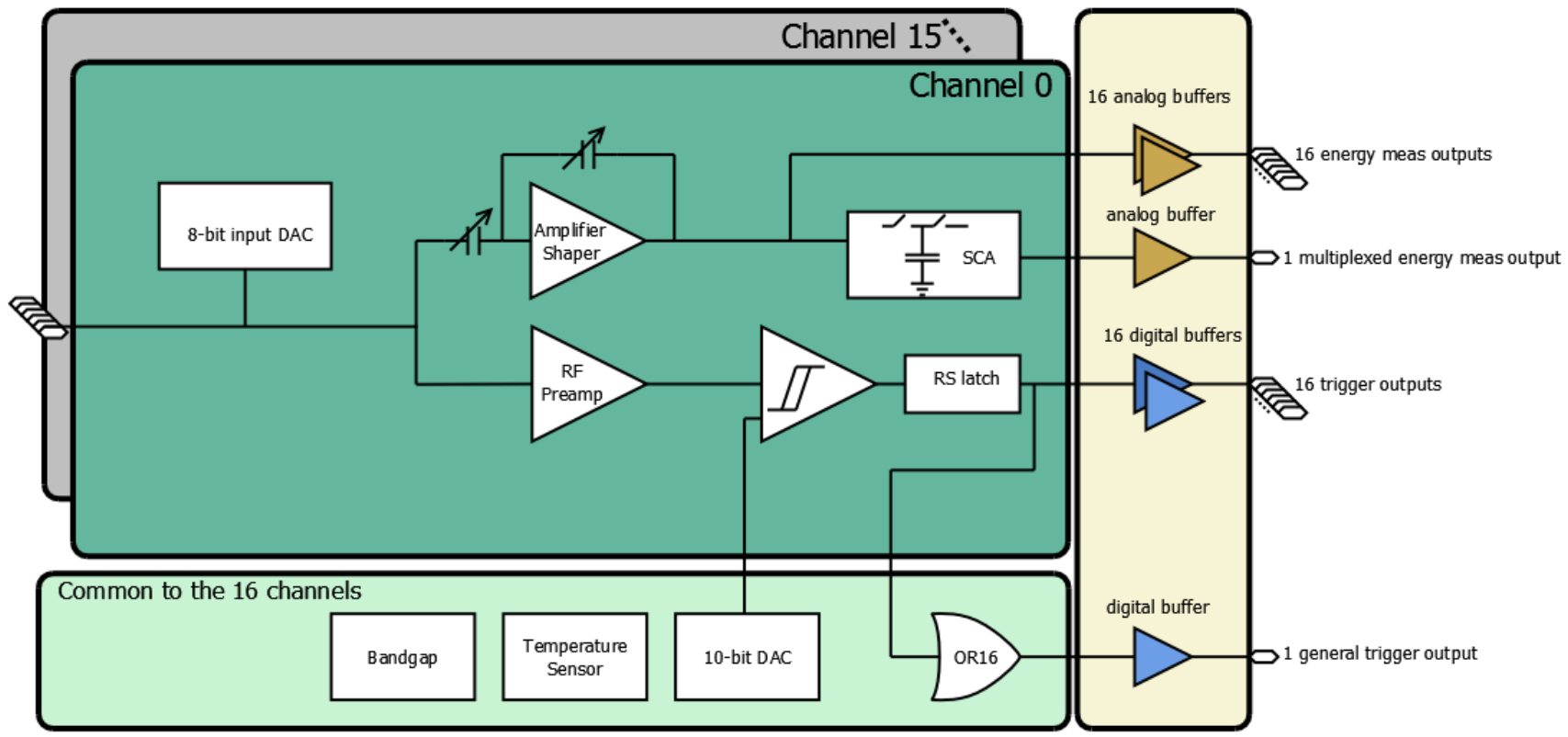}
	\caption{\label{fig:PETIROC} Block diagram of PETIROC front-end ASIC.}
\end{figure}

New prototypes of front-end electronics using SiGe technology have been proposed as shown in figures~\ref{fig:ATLAS} \&~\ref{fig:PETIROC}. The first figure presents a prototype of new preamplifier developed by Cardarelli and others~\cite{ATLAS}. This low-noise preamplifier can be fused into the current CMS front-end electronics ASIC as replacement of the actual preamplifier~\cite{FEB}. Tested on a CMS RPC, this electronics permitted a shift of the efficiency curve of \unit{460}{V} to lower values corresponding to a average charge reduction from \unit{20}{pC} to \unit{3}{pc}~\cite{PHASEIITP}.

Figure~\ref{fig:PETIROC} is a block diagram of the 16-channel PETIROC ASIC developed by OMEGA~\cite{PETIROC} for Time of Flight applications. This type of electronics is now being adapted for usage with multigap RPCs with high time resolution. The prototype offers the possibility to use 3 detection thresholds, enabling a semi-digital readout, and to read-out both strip ends. Coupled with the timing performance of multigap RPCs, a good position information along the strips can be obtained.

\begin{figure}[ht!]
	\centering
	\includegraphics[width=0.6\textwidth]{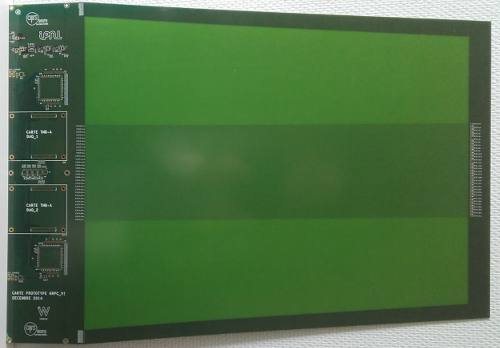}
	\caption{\label{fig:Petiroc} Picture of a strip board designed for a RE1/1 station using 16-channel PETIROC ASICs and \unit{25}{ns} TDC.}
\end{figure}

\subsection{Chamber design}
\label{subsec:design}

Choosing a better performing RPC design, with optimized electrode or gas gap thickness and various number of gaps, to combine with the previously presented front-end electronics prototypes can be investigated to further reduce the mean charge deposition per avalanche and thus the charge recombination time and the related detector dead time.

\begin{figure}[ht!]
	\centering
	\includegraphics[width=0.5\textwidth]{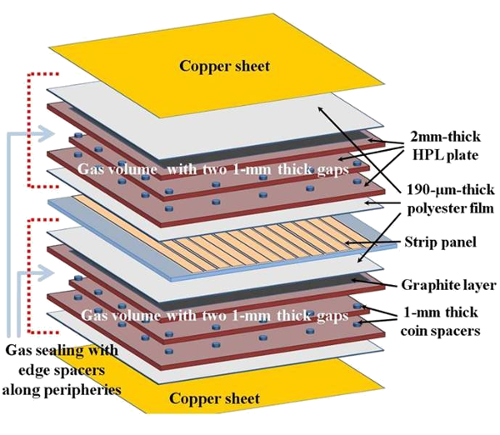}\hfill\includegraphics[width=0.45\textwidth]{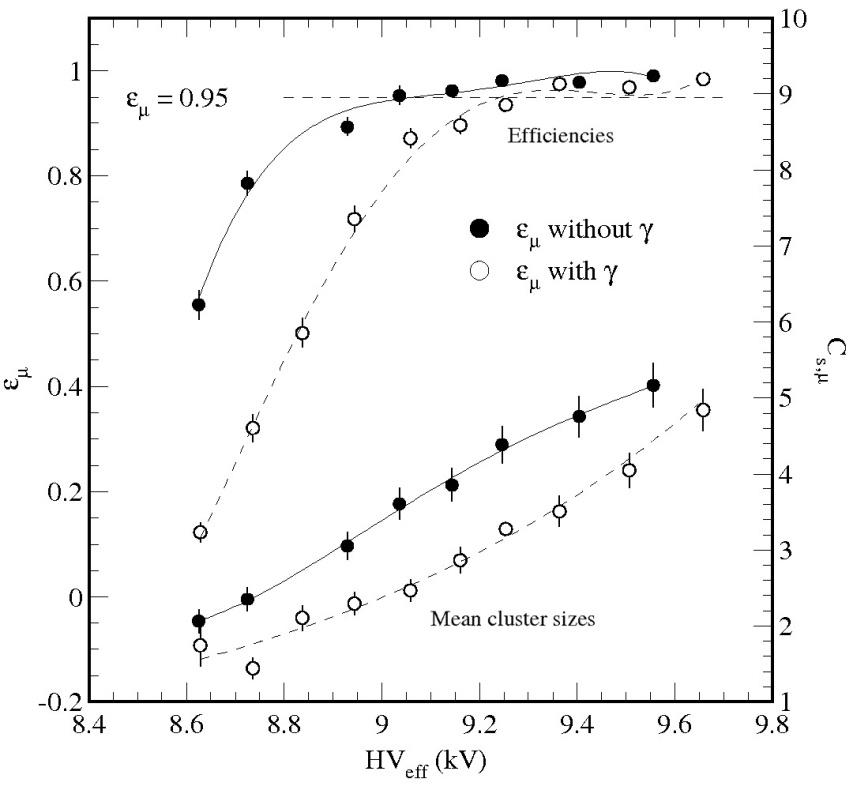}\\
	\caption{\label{fig:Kodel} Left: view of a multigap HPL RPC prototype. Right: efficiency plateau and cluster size as function of operating voltage with (open symbols) and without (solid symbols) a \unit{3}{kHz/cm\squared} $\gamma$-irradiation.}
\end{figure}

\begin{figure}[ht!]
	\centering
	\includegraphics[width=0.5\textwidth]{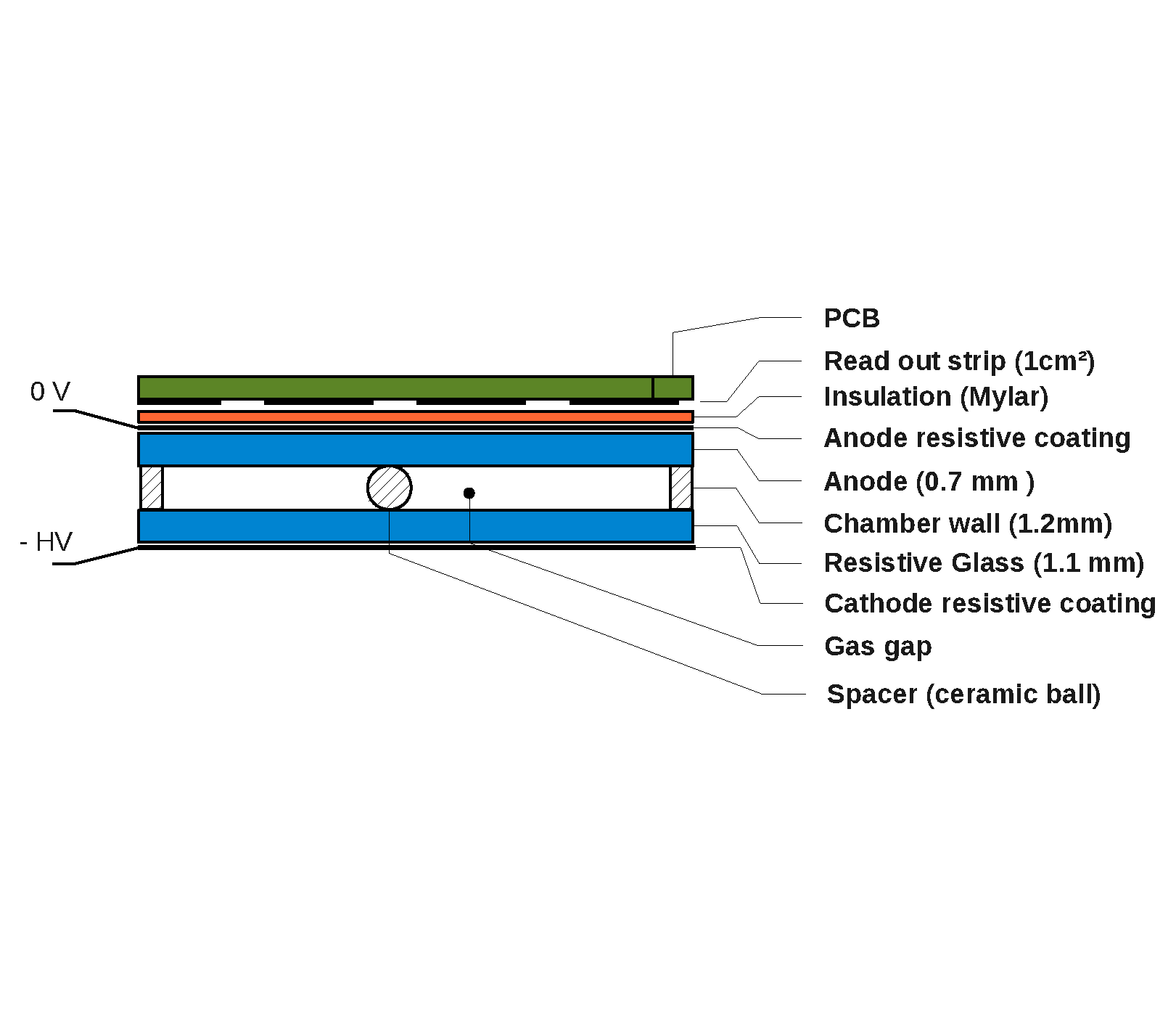}\hfill\includegraphics[width=0.45\textwidth]{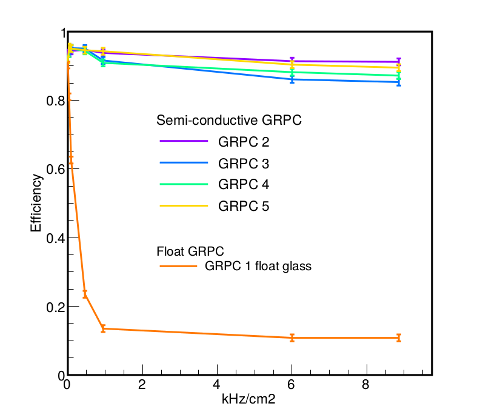}\\
	\caption{\label{fig:IPNL} Left: schematic drawing of GRPC with electrodes made of low resistive silicate glass or float glass. Right: efficiency vs rate for different RPC. The orange line corresponds to the GRPC with float glass. The semi-conductive chambers are represented with different colors.}
\end{figure}

\begin{figure}[ht!]
	\centering
	\includegraphics[width=0.5\textwidth]{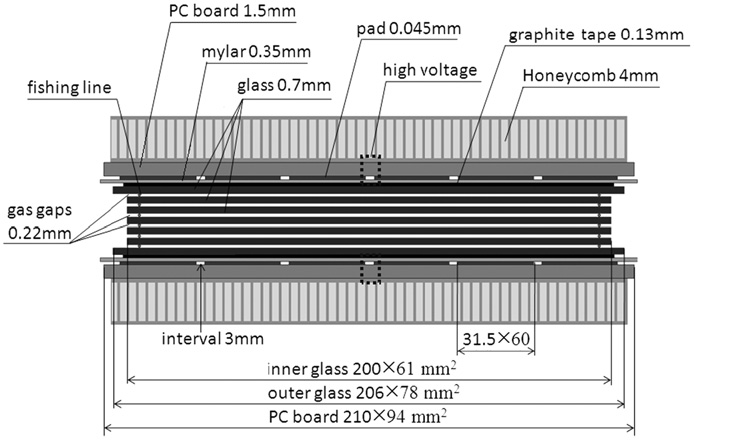}\includegraphics[width=0.5\textwidth]{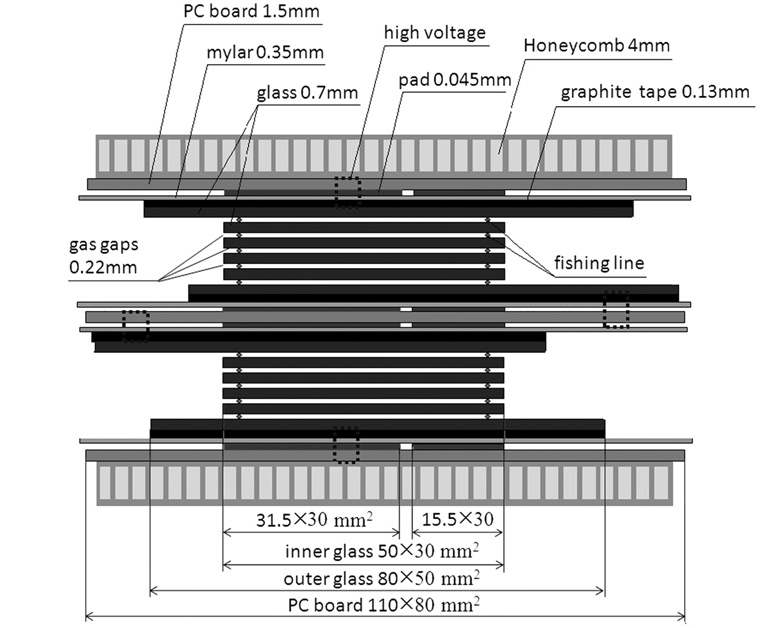}\\
	\vspace{4mm}
	\includegraphics[width=0.7\textwidth]{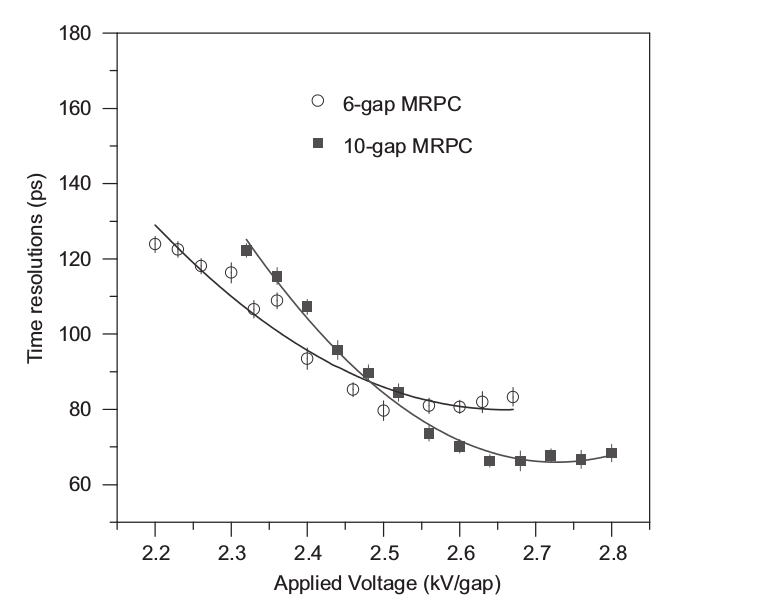}
	\caption{\label{fig:Tsinghua} Structure of 6-gap (top) and 10-gap (center) resistive plate chambers from Tsinghua University. Time resolution of 6-gap and 10-gap resistive plate chambers (bottom) as a function of high voltage at an incoming muon rate of \unit{1.2}{kHz/cm\squared}.}
\end{figure}

A bakelite based multigap solution is currently under development at KODEL~\cite{KODEL}. Figure~\ref{fig:Kodel} shows a double bi-gap HPL RPC, composed of 2 bi-gap RPCs using a design similar to that of the CMS RPC with \unit{2}{mm} electrodes and gas gaps and the results obtained with and without $\gamma$-irradiation. Using the standard CMS front-end electronics, a shift of approximately \unit{200}{V} towards higher voltages can be observed under \unit{3}{kHz/cm\squared} irradiation rate.

Combining low resistivity electrodes with low noise electronics leads to the possibility to use single gap RPCs. Such a prototype made out of LRS glass and using a similar front-end electronics as the PETIROC called HARDROC, has been proposed and and has only a limited efficiency drop compared to regular float glass for increasing particle rate~\cite{YACINE}, as shown in in figure~\ref{fig:IPNL}.

Finally, multigap prototypes made out of the same LRS glass have been made and tested with PETIROC electronics~\cite{TSINGHUA}. The very high timing precision of such RPC designs is shown figure~\ref{fig:Tsinghua}.

\section{Conclusion}
\label{sec:conclusion}

In view of the HL-LHC, the CMS RPC group is designing and testing new RPC prototypes, using new different electrode materials, geometries or electronics. The final technique adopted for the next upgrade of the CMS Muon System may combine the benefits of the different approaches. Several options for new front-end electronics are being developed, and studies are ongoing to integrate the new electronics into the CMS Trigger and DAQ system. Ageing and rate capability tests of the new prototypes are ongoing at the CERN GIF++ irradiation facility.

CMS is anticipating the installation of two new RPC stations during the upcoming third LHC Long Shutdown.


\acknowledgments

Special thanks go to Jose Repond and Miguel Morales that provided a strong basis and the material needed to construct this general CMS RPC Upgrade report.



\begin{thebibliography}{99}

\bibitem{MUONTDR}
CMS Collaboration, \emph{The CMS muon project: Technical Design Report}, \href{http://cds.cern.ch/record/343814}{CERN-LHCC-97-032}, \emph{CERN, Geneva Switzerland, LHC Experiments Committee} (1997) [CMS-TDR-3]

\bibitem{PHASEIITP}
CMS Collaboration, \emph{Technical Proposal for the Phase-II Upgrade of the CMS Detector}, \href{https://cds.cern.ch/record/2020886?ln=fr}{CERN-LHCC-2015-010}, \emph{CERN, Geneva Switzerland, LHC Experiments Committee} (2015) [LHCC-P-008]

\bibitem{CMSTP}
CMS Collaboration, \emph{CMS, the Compact Muon Solenoid : technical proposal}, \href{http://cds.cern.ch/record/290969}{CERN-LHCC-94-38}, \emph{CERN, Geneva Switzerland, LHC Experiments Committee} (1994) [CERN-LHCC-P-1]

\bibitem{HF}
M. Abbrescia et al., \emph{HF production in CMS-Resistive Plate Chambers}, \href{http://dx.doi.org/10.1016/j.nuclphysbps.2006.07.002}{\emph{Nucl. Phys. B} {\bf 158} (2006) 30-34}

\bibitem{KODEL}
K.S. Lee, \emph{Rate capability study for a four-gap phenolic RPC with a $^{137}$Cs source}, \href{http://dx.doi.org/10.1088/1748-0221/9/08/C08001}{\emph{JINST} {\bf 9} (2014) C08001}

\bibitem{YACINE}
Y. Haddad et al., \emph{High rate resistive plate chamber for LHC detector upgrades}, \href{http://dx.doi.org/10.1016/j.nima.2012.11.029}{\emph{Nucl. Instr. Meth. Phys. Res. A} {\bf 718} (2013) 424-426}

\bibitem{TSINGHUA}
J. Wang et al., \emph{Development of high-rate MRPCs for high resolution time-of-flight systems}, \href{http://dx.doi.org/10.1016/j.nima.2013.02.036}{\emph{Nucl. Instr. Meth. Phys. Res. A} {\bf 713} (2013) 40-51}

\bibitem{JOSE}
N. Johnson et al., \emph{Electronically Conductive Vanadate Glasses for Resistive Plate Chamber Particle Detectors}, \href{http://dx.doi.org/10.1111/ijag.12109}{\emph{Int J Appl Glass Sci} {\bf 6} (2015) 26-33}

\bibitem{CERAMICS}
A. Laso Garcia et al., \emph{Extreme high-rate capable timing resistive plate chambers with ceramic electrodes}, \href{http://dx.doi.org/10.1088/1748-0221/7/10/P10012}{\emph{JINST} {\bf 7} (2012) P10012}

\bibitem{MIGUEL}
M. Morales et al., \emph{Conductivity and charge depletion aging of resistive electrodes for high rate RPCs}, \href{http://dx.doi.org/10.1088/1748-0221/8/01/P01022}{\emph{JINST} {\bf 8} (2013) P01022}

\bibitem{ATLAS}
R. Cardarelli et al., \emph{Performance of RPCs and diamond detectors using a new very fast low noise preamplifier}, \href{http://dx.doi.org/10.1088/1748-0221/8/01/P01003}{\emph{JINST} {\bf 8} (2013) P01003}

\bibitem{FEB}
F. Loddo et al., \emph{New developments on front-end electronics for the CMS Resistive Plate Chambers}, \href{http://dx.doi.org/10.1016/S0168-9002(00)00980-3}{\emph{Nucl. Instr. Meth. Phys. Res. A} {\bf 456} (2000) 143-149}

\bibitem{PETIROC}
J. Fleury et al., \emph{Petiroc and Citiroc: front-end ASICs for SiPM read-out and ToF applications}, \href{http://dx.doi.org/10.1088/1748-0221/9/01/C01049}{\emph{JINST} {\bf 9} (2014) C01049}

\end{thebibliography}
\end{document}